\documentclass[a4paper,10pt]{article}
\usepackage{cite}
\usepackage[ref]{overcite}
\usepackage{graphicx}
\begin{document}

\title{\bf Does Excess Quantum Noise Exist in Spontaneous Processes?}

\author{A. Aiello, M. P. van Exter, G. Nienhuis, J. P. Woerdman \\
 {\em Huygens Laboratory, Leiden University, P.O. Box 9504,
Leiden, The Netherlands}}
\maketitle

\begin{abstract}
We investigate the role of excess quantum noise in type-II
degenerate parametric down conversion in a cavity with
non-orthogonal polarization eigenmodes. Since only two modes are
involved we are able to derive an analytical
 expression for the twin-photon generation rate  measured
outside the cavity as a function of the degree of mode
nonorthogonality. Contrary to recent claims we conclude that there
is no evidence of excess quantum noise for a parametric amplifier
working so far below threshold that spontaneous processes
dominate.
\end{abstract}

\begin{flushleft}
 \end{flushleft}
\thispagestyle{empty}
%
%
The ultimate noise limit of optical devices is set by the quantum
noise. A textbook example is formed by the laser, where phase and
intensity fluctuations have a quantum origin. These fluctuations
can be seen as the consequence of having "one noise photon in the
laser mode" \cite[p. 72]{Sieg3}. For small lasers, in particular
semiconductor lasers, this quantum noise limit is easily reached
in practical cases. More recently, there has been a large body of
work pointing at the fact that the quantum noise may be enhanced
by the so called excess noise factor or Petermann $K$ factor
\cite{Petermann,Sieg1,Grang1,Lee1,Ham2,Cheng,Emile,Deutsch}. This
excess noise may occur when the laser cavity has nonorthogonal
eigenmodes; it is then as if there are $K$ noise photons in the
laser mode. There is no doubt regarding the physical reality of
excess quantum noise; it has been verified in many experiments on
lasers with nonorthogonal longitudinal \cite{Ham2}, transverse
\cite{Cheng} and polarization \cite{Lee1,Emile} modes.

However, there is a continuing debate whether the noise
enhancement arises from an amplification of spontaneously emitted
photons by the gain medium \cite{Deutsch} or from a
cavity-enhanced single-atom decay rate \cite{Lamp1}. If the latter
interpretation applied, excess noise would also be a valid concept
(far) below the oscillation threshold of the device under
consideration. In this case excess noise could be very useful; for
instance it has been claimed that it could lead to an enhanced
generation of twin photons in Spontaneous Parametric Down
Conversion (SPDC), by placing the nonlinear crystal in an unstable
cavity (which has nonorthogonal transverse eigenmodes)
\cite{Lamp2}.

The reason that there is room for uncertainty is that all
theoretical approaches of this problem are (necessarily) model
dependent. The most common experimental realization of mode
nonorthogonality concerns the transverse modes of an unstable
cavity. However, this case is intrinsically difficult to treat:
 one deals with an infinite manifold of transverse modes
which cannot be truncated  since there is no sharp distinction
between system modes (= cavity modes) and reservoir modes (= free
space modes) \cite{Lindberg}. This has motivated us to study the
effect of excess noise on cavity-enhanced SPDC, for a case where
one can construct an exactly solvable quantum theory of mode
nonorthogonality. In fact, we use a cavity with nonorthogonal {\em
polarization} modes (instead of {\em transverse} modes). Our model
comprises (and reduces to those as particular subcases) two
theoretical models
 both of which have been
{\em experimentally} verified.
 For the type-II degenerate parametric amplifier we use
the Gardiner and Savage model [16] whose validity has been also
recently  verified by Lu and Ou [14]. For the cavity with two
non-orthogonal polarization modes we adopt the model of Bretenaker
and coworkers [6] where a large polarization $K$-factor has been
demonstrated.
 The
restriction to the polarization case does not limit the validity
of our conclusions: in earlier works it has been shown
experimentally and theoretically that the polarization $K$ factor
shows all the features of a generic $K$ factor \cite{Lee1,Emile}.
The main message of our Letter is that there is {\em no}
spontaneous emission enhancement in parametric down conversion.

SPDC constitutes a natural framework in which to study
polarization excess noise. In a type-II SPDC process, two
orthogonally polarized photons are generated. Because of  crystal
anisotropy, for a fixed angular frequency only a restricted set of
spatial directions is allowed for the emitted photons. In the
degenerate case one can achieve a single allowed direction for a
collinear emission \cite{Rubin} thus, besides imperfect
phase-matching effects,
 single transverse mode operation can be
realized. An optical cavity also allows for several resonant
longitudinal modes but the double resonance condition for SPDC
restricts this number. It can be shown \cite{Lu2} that because of
crystal birefringence, for a type-II process the double resonance
condition can only be satisfied at degenerate frequency so that
the number of allowed longitudinal modes is reduced to one.

 We consider a cavity having one perfectly reflecting
mirror at position $z=-L$, and a partially reflecting mirror at $z
= 0$, as shown in Fig. 1.
\begin{figure}
\begin{center}
\includegraphics[width=8.5truecm]{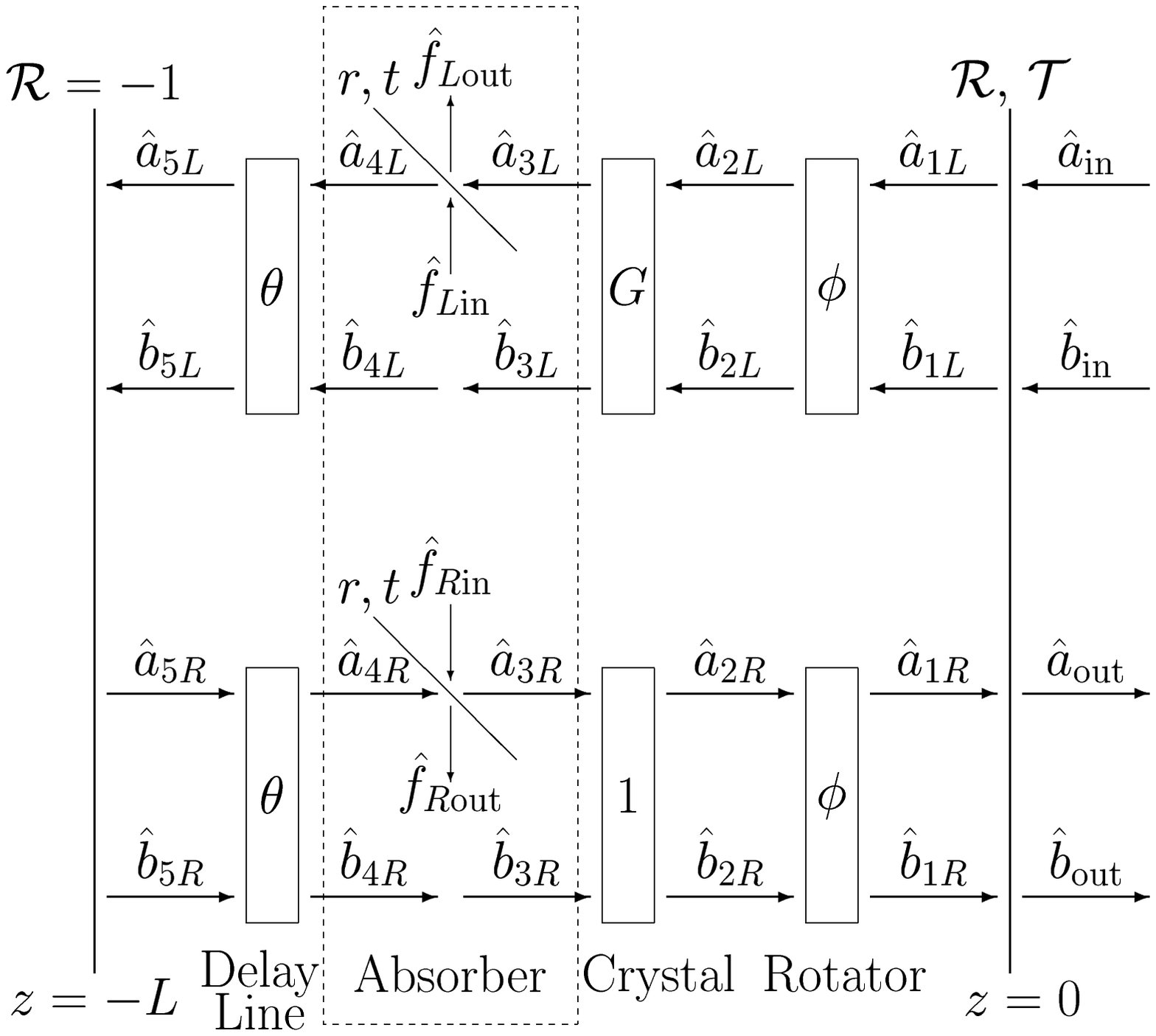}
\caption{\label{fig:1t} Schematic representation of the
degenerate-cavity parametric amplifier. Modes $a$ and $b$ have
orthogonal polarizations. The boxes indicated with $\phi$, $G$ and
$\theta$ represent the rotator, the nonlinear crystal and the
delay line, respectively. In the dashed box we show the absorber
modeled as a beam-splitter acting only on mode $a$. For the
right-travelling modes we have put $G=1$ to indicate the passive
crystal behavior when there is no phase matching. }
\end{center}
\end{figure}
 We decompose the electric field
inside the cavity into left (subscript $L$) and right (subscript
$R$) propagating waves. In degenerate type-II SPDC two
orthogonally polarized fields are generated. Thus, we have only
two modes $a$ and $b$, with
  the same angular frequency
$\omega$ and orthogonal polarization directions, parallel to the
$x$ and $y$ axis, respectively. Two other modes, $f_{L}$ and
$f_{R}$, having the same polarization as mode $a$, are introduced
in order to assure the unitarity of the model. Mode
non-orthogonality is achieved by inserting in the cavity a phase
anisotropy due to circular birefringence
 (polarization rotator) and a loss anisotropy
generated by linear dichroism (polarization dependent absorber),
following the scheme given in \cite{Emile}.

 The optical elements inside the cavity are: an absorber modeled
as a beam-splitter acting only on the $x$ polarization, a crystal
with nonlinear gain $G$ and a rotator which rotates the
polarization axes by an angle $\phi$ along $z$ axis. A delay line
has been put in front of the left mirror to account for the mode
propagation inside the cavity.
We assume that the optical elements are infinitesimally thin and
we put the  operator phases at $z = 0$ equal to zero.
It should be noticed that  because phase matching conditions can
be fulfilled only by left-travelling  or right-travelling waves
(we assume left),
 the light is amplified only  during the first half round trip
whereas it is freely propagating during the second half round
trip.
 We
characterize both lossless passive and active optical elements as
unitary devices with two input ports and two output ports related
by a scattering matrix \cite{Grang1}.

 On the right mirror
the output annihilation operators belonging to
 mode $a$, are related to the input operators on the same mode
by the transformation
\begin{equation}\label{eq:w1}
 \begin{array}{rcl}
   \hat{a}_{\mathrm{out}}     & = &
      \mathcal{T} \, \hat{a}_{1R}    +  \mathcal{R}   \,  \hat{a}_{\mathrm{in}} ,  \label{eq:1} \\
    \hat{a}_{1L}     & = &
    \mathcal{R}  \,  \hat{a}_{1R} + \mathcal{T}   \, \hat{a}_{\mathrm{in}}  \label{eq:2},
   \end{array}
    \end{equation}
where $\mathcal{R} = -\sqrt{R}$, $\mathcal{T} = i \sqrt{1-R}$ and
$0 \leq R < 1$. For mode $b$ the above relations hold if we make
everywhere the substitution $a \rightarrow b$. The effect of the
rotator on left-travelling mode operators can be represented as:
\begin{equation}
 \begin{array}{rcl}
   \hat{a}_{2L}   & = & \cos \phi  \,  \hat{a}_{1L}   +
     \sin \phi  \,   \hat{b}_{1L},  \label{eq:3} \\
   \hat{b}_{2L}   & = & -
     \sin \phi  \,   \hat{a}_{1L} +  \cos \phi  \,   \hat{b}_{1L}.  \label{eq:4}
   \end{array}
    \end{equation}
Note that we have chosen as a rotator a device antisymmetric with
respect to temporal inversion \cite{Spreeuw} (e.g., a Faraday
rotator). Then after a round trip the total rotation angle is
equal to $2 \phi$. The corresponding matrix for right-travelling
modes is obtained by substituting
 in the above formula $1 \leftrightarrow 2$ and $L
\rightarrow R$. The scattering matrix for the parametric crystal
\cite{Gardiner} is given by:
\begin{equation}\label{eq:w3}
 \begin{array}{rcl}
   \hat{a}_{3L}   & = & G  \,  \hat{a}_{2L}   +
    ( {G^2-1})^{1/2}  \,   \hat{b}_{2L}^\dag,  \label{eq:5} \\
   \hat{b}_{3L}^\dag   & = &
    ({G^2-1})^{1/2}  \,   \hat{a}_{2L} +  G  \,   \hat{b}_{2L}^\dag,  \label{eq:6}
   \end{array}
    \end{equation}
where the real gain $G$ satisfies $G>1$. The corresponding
transformation for the right-travelling modes follows after the
substitutions $3 \leftrightarrow 2$, $L \rightarrow R$ and $G =1$
(transparent medium).
Since Eqs. (\ref{eq:w3}) preserve bosonic commutation rules, for a
parametric amplifier with a classical non-depleted pump it is not
necessary to add noise from an external bath \cite{Caves} to
account for pump fluctuations. In our model only the
down-converted field is confined by the cavity, not the pump
field, therefore the cavity mode structure cannot affect the pump
beam fluctuations.
 For the absorber, which introduces losses
only for mode $a$  we have:
\begin{equation}
\begin{array}{rcl}
   \hat{a}_{4L} & = &   t  \,  \hat{a}_{3L}   + r   \,  \hat{f}_{L \mathrm{in}}
  ,  \label{eq:7} \\
   \hat{b}_{4L} & = &  \hat{b} _{3L},  \label{eq:8} \\
  \hat{f}_{L \mathrm{out}}   & = & r    \,    \hat{a}_{3L}  +  t   \,  \hat{f}_{L
  \mathrm{in}},  \label{eq:9}
\end{array}
 \end{equation}
where $r = i \sqrt{1-t^2}$  and the parameter $t$ ($0 \leq t \leq
1$) represents the ratio between field amplitudes along $x$ and
$y$ polarization directions. For right-travelling modes $4
\leftrightarrow 3$ and $L \rightarrow R$.
 For the delay line we have $\hat{a}_{5L}  =   \exp(i \theta) \,
 \hat{a}_{4L}$,
where $\theta = \omega L /c$. For right-travelling modes $5
\leftrightarrow 4$ and $L \rightarrow R$. The same relation holds
for mode $b$. Finally, on the left mirror $ \hat{a}_{5R}  =   -
\hat{a}_{5L}$
and similarly for mode $b$.\\ The equations given above can be
solved to express right-travelling mode operators in terms of
left-travelling mode operators:
%
\begin{equation}
 \begin{array}{rcl}
\left( \begin{array}{cc}
 \hat{a}_{1R} \\
 \hat{b}_{1R}
\end{array} \right) & = & \displaystyle - G
\left(\begin{array}{cc}
 C_{12}^{+}   & S_{1} \\
  -S_{1}  &  C_{12}^{-}
\end{array}\right) \left( \begin{array}{cc}
 \hat{a}_{1L} \\
 \hat{b}_{1L}
\end{array}\right) \nonumber\\
 & &  + \displaystyle \tilde{G}
\left(\begin{array}{cc} S_{2}   & - C_{21}^{+} \\
  C_{21}^{-}  & S_{2}
\end{array}\right) \left( \begin{array}{cc}
 \hat{a}_{1L}^{\dag} \\
 \hat{b}_{1L}^\dag
\end{array} \right)+ \left( \begin{array}{cc}
 \hat{f}_{a} \\
 \hat{f}_{b}
\end{array}\right), \label{eq:12}
\end{array}
 \end{equation}
%
where $\tilde{G}\equiv (G^2-1)^{1/2}$, $\gamma_{j} = \exp( 2 i
\theta)(t^2-(-1)^j)/2$, $S_{j} = \gamma_{j} \sin(2 \phi)$,
$C_{ij}^{\pm} = \gamma_i \cos(2 \phi) \pm \gamma_{j}$, ($i,j=1,2$)
and
\begin{equation}
\left( \begin{array}{c}
 \hat{f}_{a} \\
 \hat{f}_{b}
\end{array} \right)
 =   r \left(
  \begin{array}{c}
 (t \hat{f}_{L \mathrm{in}}+\hat{f}_{R \mathrm{in}})   \cos \phi  \\
  -  (t \hat{f}_{L \mathrm{in}}+\hat{f}_{R \mathrm{in}} )  \sin \phi
\end{array} \right). \label{eq:13}
\end{equation}
For a non-zero rotation angle $\phi$, the quantum noise operators
$\hat{f}_a$ and $\hat{f}_b$ do not commute:
\begin{equation}
 \begin{array}{rcl}
   \, [ \hat{f}_a, \hat{f}_a^\dag ] & = &   (1-t^4) \cos^2 \phi   , \label{eq:14}\\
   \, [ \hat{f}_b, \hat{f}_b^\dag ] & = &  (1-t^4)  \sin^2 \phi , \label{eq:15} \\
   \, [ \hat{f}_a, \hat{f}_b^\dag ] & = &   -  (1-t^4) \sin \phi \cos \phi. \label{eq:16}
\end{array}
 \end{equation}
This noise correlation disappears when the eigenmodes of the
classical cold cavity round-trip matrix ${\mathbf{M}}$
\cite{Grang1}, become orthogonal ($\phi=0$ and/or $t=1$). The
 matrix ${\mathbf{M}}$ can be obtained from Eq.
(\ref{eq:12}) by putting $G=1$ (cold cavity) and disregarding the
quantum noise term (classical). In fact  ${\mathbf{M}}$ coincides
with the matrix in the first row of Eq. (\ref{eq:12}).
 Diagonalizing the non-unitary matrix ${\mathbf{M}}$ we find for
the Petermann $K$ factor
\begin{equation}
K_< = \frac{(1-t^2)^2}{(1-t^2)^2 - (1+t^2)^2 \sin^2 2\phi},
\label{eq:17}
\end{equation}
when $t<t_c$ (locked regime \cite{Lee1}), and
\begin{equation}
K_> = \frac{(1+t^2)^2 \sin^2 2\phi}{ (1+t^2)^2 \sin^2 2\phi -
(1-t^2)^2}, \label{eq:18}
\end{equation}
for $t>t_c$ (unlocked regime), where $t_c (\phi) = \sqrt{(1-|\sin
2 \phi|)/(1 + |\sin 2\phi|)}$.
 Apart from notation these results
agree with earlier works \cite{Lee1,Emile}. We conclude that our
model correctly accounts for non-orthogonal polarization modes.
\begin{figure}
\begin{center}
\includegraphics[width=6.5truecm]{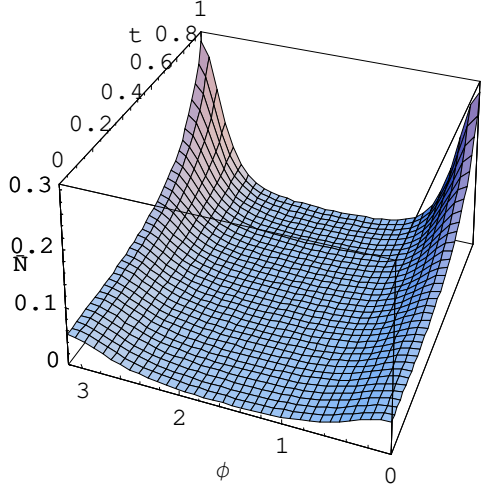}
\caption{\label{fig:2}Plot of the total average photon number
$\bar{N} \equiv \bar{n}_a+\bar{n}_b$ of the sub-threshold OPO,
calculated at resonance, as function of the absorber transmission
$t$ and of the rotator angle $\phi$. The values of the other
parameters are: $G = 1.01$, $R = 0.2$. For $t=0$ photons in mode
$a$ are fully absorbed and the
  residual value of $\bar{N}$ is due to
 contribution of only mode $b$.}
\end{center}
\end{figure}

 The next step is to calculate the SPDC rate
and to study how it depends on the "non-orthogonality parameters"
$t$ and $\phi$. Equations (\ref{eq:w1}) together with Eqs.
(\ref{eq:12}) can be straightforwardly solved to express "out"
operators in terms of "in" operators. The resulting expressions
are very cumbersome and it is not useful to write them explicitly.
Their general form is
\begin{equation}
\hat{a}_{\mathrm{out}} =  \sum_{\alpha = a,b} \left( M_{1\alpha}
\hat{\alpha}_{\mathrm{in}} + M_{2 \alpha}
\hat{\alpha}_{\mathrm{in}}^\dag + M_{3\alpha} \hat{f}_{\alpha } +
M_{4\alpha} \hat{f}_{\alpha }^\dag \right), \label{eq:19}
\end{equation}
and similarly for mode $b$, where $M_{i \alpha}$ are complicated
functions of $t, \phi, G, R$ and $L$. From the above formula we
calculate the average photon number emitted on modes $a$ and $b$:
\begin{equation}
\bar{n}_{\alpha} = \langle \hat{\alpha}_\mathrm{out}^\dag
\hat{\alpha}_\mathrm{out}
  \rangle_\mathrm{vac}, \qquad (\alpha=a,b), \label{eq:20}
\end{equation}
where the subscript "vac" indicates that the quantum expectation
value is calculated for incoming vacuum field. When both absorber
and rotator are switched off (orthogonal modes case)
$\bar{n}_{a}=\bar{n}_{b}\equiv \bar{n}$, where
\begin{equation}
\bar{n} = (G^2-1) \left[ \frac{1-R}{1-2G\sqrt{R} \cos(2 \omega
L/c) + R } \right]^2\label{eq:21}.
\end{equation}
The term inside the square brackets, when calculated for $G=1$,
coincides with the  spontaneous emission modification factor $F$
\cite{Ueda}, but in our case it is quadratic because of
nonlinearity \cite{Aiello}. At resonance ($L = m \pi c / \omega$,
with $m$ integer), a divergence appears for $\bar{n}$ when $G =
(1+R)/(2 \sqrt{R})>1$, corresponding to the threshold of
oscillation \cite{Reynaud}. However, we are interested only in the
sub-threshold case where  a privileged lasing mode is not
selected.
\begin{figure}
\begin{center}
\includegraphics[angle=0,width=8truecm]{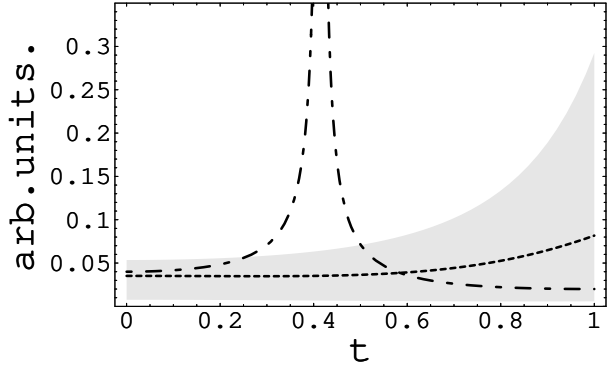}
\caption{\label{fig:3} Dotted-dashed line: "geometrical" Petermann
$K$ factor, given by Eqs. (\ref{eq:17}-\ref{eq:18}) for a cavity
without crystal, calculated for $\phi = \pi/8$, as a function of
the absorber transmission $t$. The value of $K$ diverges for $t
\rightarrow t_c(\phi = \pi / 8) \simeq 0.41$.  Dashed line: total
average photon number $\bar{N}$ calculated  at resonance and $\phi
= \pi/8$. The values of the other parameters are $G = 1.01$, $R =
0.2$, corresponding to a sub-threshold OPO. The grey band
represents all possible values of $\bar{N}$ for non-orthogonal
modes. Note that $\bar{N}$ is not enhanced for $t=t_c$. }
\end{center}
\end{figure}
We report in Fig. 2  the total average photon number $\bar{N}
\equiv \bar{n}_b+\bar{n}_a$, evaluated at resonance, as function
of the absorber transmission coefficient $t$ and of the rotation
angle $\phi$ due to the rotator.  The nonlinear gain $G$ and the
output mirror reflectivity $R$ have been chosen as $G = 1.01$, $R
= 0.2$, so that sub-threshold operation is achieved.

From Fig. 2 it is clear that the local maxima of $\bar{N}$, for
$t$ variable, are located on the curve $\phi =0$ which corresponds
to a cavity with orthogonal
 modes. This curve constitutes the upper boundary of the grey band
shown in Fig. 3. The other points in the grey band
  represent all  possible values of $\bar{N}$, calculated with the same parameters as in Fig. 2,
   for cavities with
  non-orthogonal modes. All these points are below the curve
  corresponding to orthogonal modes; so we do {\em not} find any
  twin-photon
  enhancement under these conditions.

 This may be compared with the behavior of the  "geometrical" $K$ factor,  as given by
 Eqs. (\ref{eq:17}-\ref{eq:18}), that is the $K$ factor
 which appears in a "cold cavity" (i.e., without gain) as simple consequence
of mode nonorthogonality. Fig. 3 shows the behavior of this $K$
factor with respect to $\bar{N}$, as function of the absorber
transmission $t$. Both $K$ and $\bar{N}$  are evaluated for $\phi
= \pi/8$. The values of the other parameters for $\bar{N}$ are the
same as in Fig. 2. From a geometrical point of view, when $t=t_c$
the cavity eigenmodes become parallel and the corresponding $K$
factor diverges. In Fig. 3 this resonant behavior of $K$, when $t$
approaches $t_c$, is evident, but at the same time there is no
signature of a critical
 behavior of $\bar{N}$. Therefore we conclude that for a
sub-threshold OPO, the total average photon number $\bar{N}$ does
not depend on $K$.

 In conclusion we have shown that there is no excess quantum noise
enhancement in type-II SPDC. On the contrary, the use of a cavity
with nonorthogonal (instead of orthogonal) eigenmodes leads to a
reduction of the twin photon generation rate. Excess quantum noise
must therefore be exclusively ascribed to amplification of
spontaneously emitted photons; the spontaneous emission process
itself is not affected. Excess quantum noise becomes effective
only very close to threshold when one of the cavity eigenmodes is
"selected" as the lasing mode which dominates over the other modes
\cite{Haus}.

\section*{Acknowledgments}
We acknowledge support from the EU under the IST-ATESIT contract.
This project is also part of the program of FOM.


\begin{thebibliography}{aa99}
\bibitem{Sieg3}
A.~E.Siegman,
  \emph{{Lasers}} ({{Univ. Sci. Books, Mill
  Valley, CA}},    {1996}).


\bibitem{Grang1}
  { {P.}~  {Grangier}}   {and}
    { {J.-P.}   {Poizat}},
     {Eur. Phys. J. D} \textbf{   {1}},
     {97} ({1998});
      \emph{{ibid.}} \textbf{   {7}},
     {99} ({1999}).

\bibitem{Lee1}
  { {A.~M.}   {van~der Lee}} \emph{{et al.}},
 {Phys.\ Rev. Lett.}
  \textbf{   {79}},    {4357} ({1997});
%
  { {A.~M.}   {van~der Lee}} \emph{{et al.}},
   {Phys.\ Rev. A}
  \textbf{   {61}},    {033812}
  ({2000}).

\bibitem{Ham2}
  { {W.~A.}   {Hamel}}   {and}
    { {J.~P.}   {Woerdman}},
     {Phys.\ Rev. Lett.} \textbf{   {64}},
     {1506} ({1990}).

\bibitem{Cheng}
  { {Y.-J.}   {Cheng}},
    { {C.~G.}   {Fanning}},
    {and}   { {A.~E.}
    {Siegman}},    {Phys.\ Rev. Lett.}
  \textbf{   {77}},    {627} ({1996});
%
  { {M.~A.}   {van Eijkelenborg}},
    { {{\AA }.~M.}   {Lindberg}},
    { {M.~S.}   {Thijssen}},
    {and}   { {J.~P.}
    {Woerdman}},    \emph{{ibid.}}
  \textbf{   {77}},    {4314} ({1996});
%
  { {A.~M.}   {van~der Lee}} \emph{{et al.}},
     \emph{{ibid.}} \textbf{   {85}},
     {4711} ({2000}).

\bibitem{Emile}
  { {O.}~  {Emile}},
    { {M.}~  {Brunel}},
    { {A.~L.}   {Floch}},   {and}
    { {F.}~  {Bretenaker}},
     {Europhys. Lett.} \textbf{   {43}},
     {153} ({1998}).

\bibitem{Deutsch}
  { {I.~H.}   {Deutsch}},
    { {J.~C.}   {Garrison}},
    {and}   { {E.~M.}
    {Wright}},    {J. Opt. Soc. Am. B}
  \textbf{   {8}},    {1244} ({1991});
%
  { {P.}~  {Goldberg}},
    { {P.~W.}   {Milonni}},
    {and}   { {B.}~  {Sundaram}},
     {Phys.\ Rev. A} \textbf{   {44}},
     {1969} ({1991}).

\bibitem{Petermann}
  { {K.}~  {Petermann}},
     {IEEE J. Quant. Electron.} \textbf{   {QE-15}},
     {566} ({1979}).

\bibitem{Sieg1}
  { {A.~E.}   {Siegman}},
     {Phys.\ Rev. A} \textbf{   {39}},
     {1253} ({1989});
%
    \emph{{ibid.}} \textbf{   {39}},
     {1264} ({1989}).

\bibitem{Lamp1}
  { {C.}~  {Lamprecht}}   {and}
    { {H.}~  {Ritsch}},
     {Phys. Rev. Lett.} \textbf{   {82}},
     {3787} ({1999});
%
     {Phys. Rev. A} \textbf{   {65}},
     {023803} ({2002});
%
  { {S.~A.}   {Brown}}   {and}
    { {B.~J.}   {Dalton}}
  ({2001}),  {quant-ph/0107039}.

\bibitem{Lamp2}
  { {C.}~  {Lamprecht}},
    { {M.~K.}   {Olsen}},
    { {M.}~  {Collett}},   {and}
    { {H.}~  {Ritsch}},
     {Phys. Rev. A} \textbf{   {64}},
     {033811} ({2001}).

\bibitem{Lindberg}
  { {A.~M.}   {Lindberg}},
    { {K.}   {Joosten}},
    { {G.}~  {Nienhuis}},   {and}
    { {J.~P.}   {Woerdman}},
     {Opt. Commun.} \textbf{   {153}},
     {55} ({1998}).

\bibitem{Rubin}
  { {M.~H.}   {Rubin}},
     {Phys. Rev. A} \textbf{   {54}},
     {5349} ({1996}).

\bibitem{Lu2}
  { {Y.~J.}   {Lu}}   {and}
    { {Z.~Y.}   {Ou}},
     {Phys. Rev. A} \textbf{   {62}},
     {033804} ({2000}).


\bibitem{Spreeuw}
  { {R.~J.~C.}   {Spreeuw}},
    { {M.~W.}   {Beijersbergen}},
    {and}   { {J.~P.}
    {Woerdman}},    {Phys. Rev. A}
  \textbf{   {45}},    {1213} ({1992}).

%

\bibitem{Gardiner}
  C. W. Gardiner and C. M. Savage,
     Opt. Commun. A \textbf{50},
     173 (1984).

\bibitem{Caves}
C. M. Caves, Phys. Rev. D \textbf{26}, 1817 (1982).

\bibitem{Ueda}
  { {M.}~  {Ueda}}   {and}
    { {N.}~  {Imoto}},
     {Phys. Rev. A} \textbf{   {50}},
     {89} ({1994}).

\bibitem{Aiello}
  { {A.}~  {Aiello}},
    { {D.}~  {Fargion}},   {and}
    { {E.}~  {Cianci}},
     {Phys. Rev. A} \textbf{   {58}},
     {2446} ({1998}).

\bibitem{Reynaud}
  { {S.}~  {Reynaud}},
    { {A.}~  {Heidmann}},
    { {E.}~  {Giacobino}},
    {and}   { {C.}~  {Fabre}},
     {Progress in Optics} \textbf{   {XXX}}
  ({1992}),    {edited by E. Wolf (North Holland,
  Amsterdam)}.


\bibitem{Haus}
  {H. A. Haus and S. K. Kawakami},
     {IEEE J. Quant. Electron.} \textbf{   {QE-21}},
     {63} ({1985}).
\end{thebibliography}
\end{document}